\documentclass{WileyMSP-template}

\usepackage{amsmath}
\usepackage{amssymb}
\usepackage{amsthm}
\usepackage{mathrsfs}
\usepackage{amsbsy}
\usepackage{amstext}
\usepackage{cite}

\usepackage{xr-hyper}
\usepackage{hyperref}
\hypersetup{colorlinks,breaklinks,
	urlcolor=[rgb]{0,0,0.64},
	linkcolor=[rgb]{0,0,0.64},
	citecolor=[rgb]{0,0,0.64},
	filecolor=[rgb]{0,0,0.64}}

\makeatletter
\newcommand*{\addFileDependency}[1]{
	\typeout{(#1)}
	\@addtofilelist{#1}
	\IfFileExists{#1}{}{\typeout{No file #1.}}
}
\makeatother

\newcommand*{\myexternaldocument}[1]{%
	\externaldocument{#1}%
	\addFileDependency{#1.tex}%
	\addFileDependency{#1.aux}%
}

\myexternaldocument{supp}

\begin{document}

\pagestyle{fancy}
\rhead{\includegraphics[width=2.5cm]{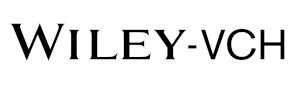}}

\title{Nonresonant Raman control of ferroelectric polarization}

\maketitle


\author{Jiaojian Shi*}
\author{Christian Heide}
\author{Haowei Xu}
\author{Yuejun Shen}
\author{Meredith Henstridge}
\author{Isabel Sedwick}
\author{Anudeep Mangu}
\author{Xinyue Peng}
\author{Shangjie Zhang}
\author{Mariano Trigo}
\author{Tony F. Heinz}
\author{Ju Li}
\author{Keith A. Nelson}
\author{Edoardo Baldini}
\author{Jian Zhou}
\author{Shambhu Ghimire}
\author{David A. Reis}
\author{Aaron M. Lindenberg*}



\begin{affiliations}
J. Shi, Y. Shen, A. Mangu, A. M. Lindenberg \\
Department of Materials Science and Engineering, Stanford University, Stanford, CA
94305, USA\\
Email Address: jiaojian@stanford.edu; aaronl@stanford.edu

J. Shi, M. Trigo, D. A. Reis, A. M. Lindenberg \\
Stanford Institute for Materials and Energy Sciences, SLAC National Accelerator Laboratory, Menlo Park, CA 94025, USA\\

J. Shi, I. Sedwick\\
Department of Chemistry, University of Washington, Seattle, WA 98195, USA\\

C. Heide\\
Department of Applied Physics, Stanford University, Stanford, CA 94305, USA\\

C. Heide, M. Trigo, T. F. Heinz, S. Ghimire, D. A. Reis, A. M. Lindenberg\\
Stanford PULSE Institute, SLAC National Accelerator Laboratory, Menlo Park, CA 94025, USA\\

H. Xu, J. Li\\
Department of Nuclear Science and Engineering, Massachusetts Institute of Technology, Cam-
bridge, MA 02139, USA\\

M. Henstridge\\
Laser Science and Technology, SLAC Linear Accelerator Laboratory, Menlo Park, CA
94025, USA\\

X. Peng, S. Zhang, E. Baldini\\
Department of Physics, Center for Complex Quantum Systems, The University of Texas at
Austin, Austin, TX 78712, USA

T. F. Heinz\\
E. L. Ginzton Laboratory, Stanford, CA 94305, USA\\

J. Li\\
Department of Materials Science and Engineering, Massachusetts Institute of Technology, Cam-
bridge, MA 02139, USA

K. A. Nelson\\
Department of Chemistry, Massachusetts Institute of Technology, Cambridge, MA 02139, USA

J. Zhou\\
Center for Alloy Innovation and Design, State Key Laboratory for Mechanical Behavior of Materials, Xi’an Jiaotong University, Xi’an 710049, CN\\

\end{affiliations}


\keywords{Impulsive stimulated Raman scattering, Phase transition, Ferroelectricity}

\begin{abstract}

Important advances have recently been made in the search for materials with complex multi-phase landscapes that host photoinduced metastable collective states with exotic functionalities. In almost all cases so far, the desired phases are accessed by exploiting light-matter interactions via the imaginary part of the dielectric function through above-bandgap or resonant mode excitation. Nonresonant Raman excitation of coherent modes has been experimentally observed and proposed for dynamic material control, but the resulting atomic excursion has been limited to perturbative levels. Here, this challenge is overcome by employing nonresonant ultrashort pulses with low photon energies well below the bandgap. Using mid-infrared pulses, ferroelectric reversal is induced in lithium niobate, and the large-amplitude mode displacements are characterized through femtosecond stimulated Raman scattering and second harmonic generation. This approach, validated by first-principle calculations, defines a novel method for synthesizing hidden phases with unique functional properties and manipulating complex energy landscapes at reduced energy consumption and ultrafast speeds.

\end{abstract}

\section{Introduction}

\begin{figure}
  \includegraphics[width=\linewidth]{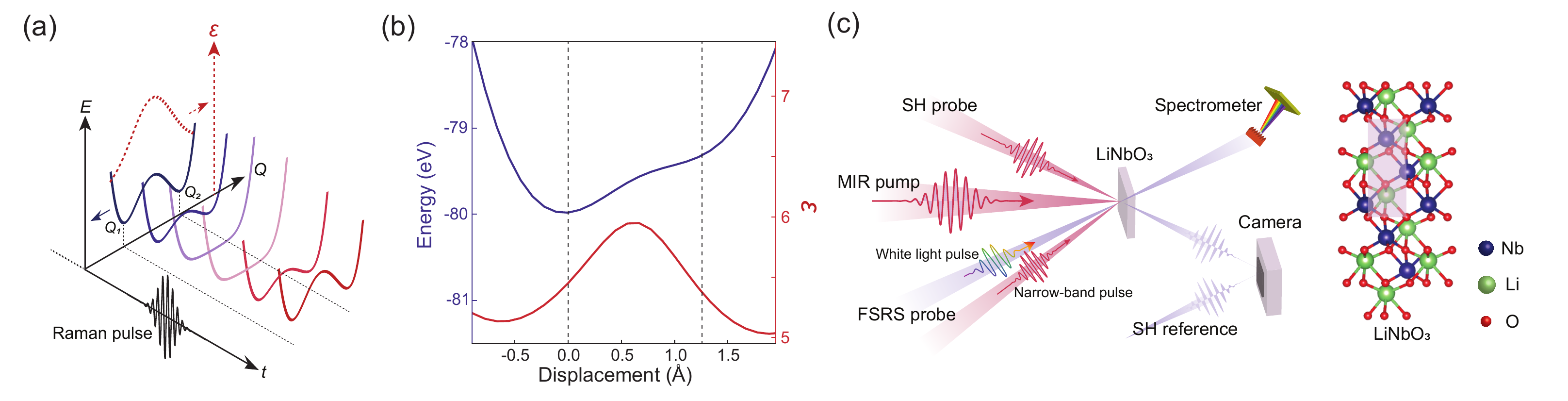}
  \caption{Nonresonant Raman control of the ferroelectric potential and experimental setup. a) The equilibrium double-well potential, which characterizes ferroelectric phase transitions in LiNbO$_3$, can be modified by applying nonresonant Raman pulses. These pulses interact with a displacement($Q$)-dependent polarizability $\varepsilon$. For the case when the dielectric constant is a maximum at the saddle point between $Q_1$ and $Q_2$, this interaction gives a force towards the saddle point when subjected to an electric field pulse, thereby altering the potential energy landscape. b) Theoretical calculation of potential energy and static dielectric constant of LiNbO$_3$ as a function of displacement along the soft mode coordinate. c) Setup schematic of MIR-pump FSRS-probe, SHG-probe, and phase-sensitive SHG-probe measurements in LiNbO$_3$. The FSRS probe uses a white light pulse and a narrow-band pulse to spectrally monitor Raman fingerprints. In the MIR-pump phase-sensitive SHG-probe measurements, time-dependent SHG from the MIR-excited sample interferes with a reference SHG from an unexcited sample and is measured with a camera.}
  \label{fig:1}
\end{figure}

In the last decades, there have been enormous efforts investigating light-induced phase transitions by resonantly exciting one degree of freedom or its coupled motions, mediated by the imaginary part of the dielectric function~\cite{Kampfrath2013,Rodenas2019,Valmispild2024,Kwaaitaal2024}.
This approach typically involves excitation of phonons~\cite{Fausti2011,Sie2019,Nova2019,Li2019,Juraschek2021}, charge~\cite{Baierl2016,Jimenez-Galan2020,Baum2007,Vaskivskyi2024,Yang2019,Liu2012,Zhang2016,Stoica2019,Huang2022}, and spin~\cite{Schlauderer2019}. An alternative method is impulsive stimulated Raman scattering (ISRS),  in which a nonresonant short pulse excites coherent modes~\cite{Shen1965,Silvestri1985,Yan1985,Ruhman1987,Merlin1997,Liu1994}. Following its success as a spectroscopic tool~\cite{Raman1928, Dhar1994}, proposals to use ISRS to drive phase transitions were put forward, using single pulses or timed shaped pulse trains~\cite{Weiner1990,Smith1996,Hiller1996} in condensed matter~\cite{Fahy1994} and molecular systems~\cite{Nelson1991}, but have never been experimentally demonstrated. This Raman force perspective can be envisioned as a dynamic shift in the potential energy landscape, where a laser field alters the energy profile through its interaction with displacement-dependent polarizability, allowing for energy transfer mediated by the real part of the dielectric function. As a prototypical example, this could represent a means for ultrafast reconfiguration of the ferroelectric polarization in a double well potential, as shown in Figure~\ref{fig:1}a. In an one-dimensional approximation, the Raman force scales as $(\partial\varepsilon/\partial Q)E^2$, where $\varepsilon$ is the real part of the dielectric function, $Q$ is the phonon coordinate, and $E$ represents the incident laser field. For typical differential polarizabilities~\cite{Cardona2005} and incident fields of a few V/nm, the resulting forces range from 1 to 10~nN, theoretically inducing displacements on the order of angstroms within 100~fs.

Despite the great potential the Raman force holds for manipulating material structures, it has not been achieved. The atomic excursion driven by this approach via near-infrared or visible laser excitation has been perturbative, and these prospects have been hindered by the concomitant excitation of carriers and subsequent heating-induced sample damage~\cite{Dastrup2017}. Here, we demonstrate that it is possible to overcome this challenge by employing nonresonant ultrashort pulses with low photon energies significantly below the bandgap. We achieve this in a prototypical ferroelectric, lithium niobate (LiNbO$_3$), using mid-infrared (MIR) pulse excitation and concurrently monitoring the lattice dynamics using femtosecond stimulated Raman scattering and second harmonic generation (SHG). Large-amplitude displacements of the ferroelectric soft mode, driven by nonresonant Raman excitation, reverse both the sign of the Raman tensor coefficients and the second harmonic phase, indicating ferroelectric switching and excitation of angstrom-level displacements.

LiNbO$_3$, with its large 3.8~eV bandgap, was chosen as a model ferroelectric for demonstrating fully reversible ferroelectric switching via Raman excitation and capturing ultrafast reversal dynamics beyond prior laser-writing studies~\cite{Yang2008,Kosters2009,Xu2022}. Supplementary Figure~3 shows Fourier-transform infrared transmission measurements confirming high transparency within the excitation frequency range. Figure~\ref{fig:1}b shows theoretical calculations of potential energy and dielectric constant as functions of atomic displacement along the soft mode coordinate. The potential energy surface reveals two local minima corresponding to the ferroelectric phases. In the case where the dielectric function peaks at the saddle point between a double-well potential, an applied electric field can trigger a Raman force toward the other well. The experimental setup, illustrated in Figure~\ref{fig:1}c, employs MIR fields centered at 5~$\mu$m to initiate phonon coherence through nonresonant Raman excitation. Two probe pulses are utilized for FSRS detection~\cite{Kukura2007}: a white-light continuum pulse with a 50-fs FWHM and a spectral bandwidth of approximately 100~THz centered at 800~nm, as well as a second pulse with a narrow spectral bandwidth ($<$~1~THz) centered at 800~nm and a duration of approximately 3~ps. Nonlinear interaction between the pulses and a Raman-active mode in the material induces oscillation and generates Raman sidebands, which interfere with white-light pulse and are detected by a spectrometer. FSRS has the merit of allowing for the detection of all Raman modes in a single probing event and the observation of rapid changes in the Raman tensor~\cite{Henstridge2022}. SHG and phase-sensitive SHG were also used to detect symmetry changes and identify the transient polarization~\cite{Mankowsky2017}.

\section{Large Ferroelectric Displacements in Transmission Measurements}

\begin{figure}
  \includegraphics[width=\linewidth]{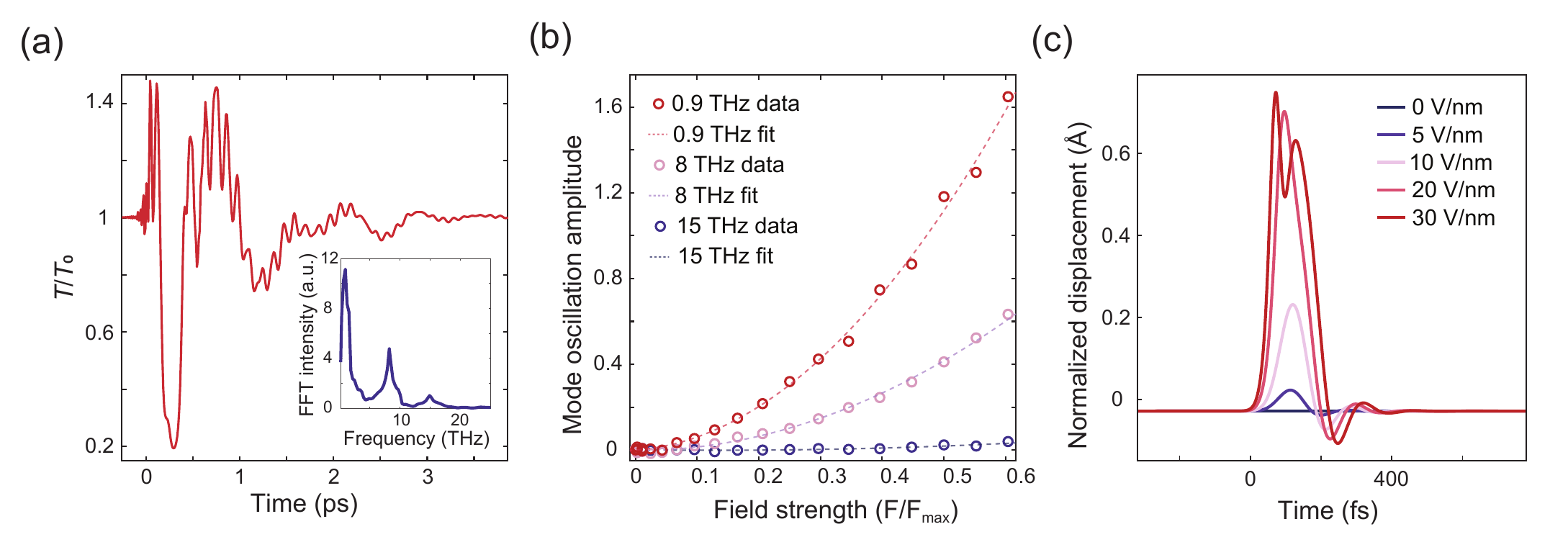}
  \caption{Large-amplitude MIR-driven phonon oscillations in LiNbO$_3$. The changes in white-light transmission over time resulting from MIR excitation reveal the presence of coherent phonon oscillations at approximately 0.9~THz, 8~THz, and 15~THz, as illustrated in the inset. b) The amplitudes of the three phonon modes are derived from the Fourier transform of responses after the delay of about 400~fs. They show quadratic dependence on the MIR field strength, which supports their ISRS origin. c) Theoretical simulation of time-dependent atomic displacement under MIR pulse excitation at increasing field strengths.}
  \label{fig:2}
\end{figure}

We first verified that nonresonant Raman scattering with MIR pulses can induce large displacements. Figure~\ref{fig:2}a presents MIR-pump transient transmission changes in LiNbO$_3$. The oscillatory features at positive time delays have frequencies of approximately 0.9, 8, and 15~THz, corresponding to the phonon polariton and $A_1$ modes~\cite{Henstridge2022} (detailed mode assignments and walk-off estimation are provided in Supplementary Note~1). The observation of multiple mode excitation confirms that nonresonant Raman can simultaneously excite various modes, but here with large amplitude, associated with significant changes in the transmitted white-light probe intensity at the level of ten's of percent. These phonon oscillation amplitudes scale quadratically with the incident MIR field strength, as shown in Figure~\ref{fig:2}b, which supports their nonresonant Raman origin. The response shows no saturation behavior up to applied fields of the order of 5~V/nm, above which abrupt sample damage occurs. This contrasts with the saturation observed with 800-nm excitation, as reported in the literature~\cite{Dastrup2017} and confirmed in our experiments shown in Supplementary Figure~5. We note that the extracted phonon oscillation, obtained from the Fourier transform of oscillations after the $\sim$~400~fs delay time in Figure~\ref{fig:2}b, should exhibit minimal nonlinearity. Nonlinearity primarily arises during the initial oscillation cycle related to ferroelectric reversal and is not captured in this averaged analysis at later time delays. The substantial transmission drop of 80\% at time zero, as shown in Figure~\ref{fig:2}b, is primarily due to MIR-induced band edge shifts~\cite{Ghimire2011} (see Supplementary Note~2 for details). Simulations with similar field strengths shown in Figure~\ref{fig:2}c (details are included in Supplementary Note~3) predict displacements up to 0.8~\AA, corresponding to a transient transition across the saddle point toward another ferroelectric phase.

\section{Ferroelectric Reversal in FSRS Measurements}

To explore the direct linkage between these effects and the ferroelectric phase transition, we employed FSRS, SHG, and phase-sensitive SHG as diagnostic tools. The FSRS spectrum of LiNbO$_3$ at equilibrium is depicted in Figure~\ref{fig:3}a with mode assignments included in Supplementary Figure~4. Figure~\ref{fig:3}b shows the evolution of the FSRS spectrum as the delay between the MIR and the FSRS pulse pair is varied. The time-resolved FSRS spectrum reveals several key features: (1) wiggles before time zero, (2) peak shifts between delay zero and $\sim$~150~fs, (3) a decrease in peak intensity and sign flip at delay around $250-350$~fs, and (4) periodic peak shifts at later delays corresponding to phonon oscillation cycles. Detailed explanations of features (1), (2), and (4) connected to the nonlinear wave mixing effect are described in Supplementary Note~4. We focus here on the feature (3), where distinct responses at a $\sim$~300~fs delay include a notable decrease in FSRS peak intensity, peak disappearance, and a transient sign flip around $250-350$~fs delay. This sign flip is further illustrated in the line-cut plot Figure~\ref{fig:3}c and d, where the FSRS spectrum and peak intensity versus delay time exhibit behavior similar to resonant phonon excitation~\cite{Henstridge2022}, suggesting that nonresonant Raman excitation induces a comparable level of distortion towards ferroelectric switching as the resonant excitation approach~\cite{Mankowsky2017,Henstridge2022}. Additional calculations in Supplementary Note~5 reproduce key FSRS spectrum features by incorporating Raman tensor changes, nonlinear wave mixing processes, and phonon-polariton oscillations in the model, supporting the observed reversal in Raman tensor coefficients and associated ferroelectric reversal.

\section{Ferroelectric Reversal in SHG Measurements}


Figure~\ref{fig:4}a presents results from the MIR-pump SHG-intensity-probe experiment at varying excitation fluences. MIR excitation at a fluence of $\sim$~440~mJ/cm$^2$ reduces the SHG by approximately 93\% at the delay time of 200~fs, followed by transient recovery at $\sim$~280~fs and relaxation at $\sim$~500~fs. At the delay of about 1.1~ps, the SHG is transiently enhanced, corresponding to the excitation of the 0.9~THz phonon-polariton mode, before fully returning to its equilibrium value. Detailed fluence-dependent scans are included in Supplementary Figure~9, along with an extended averaged trace in Supplementary Figure~11. We note that although the initial SHG suppression is influenced by the attenuation of the probe beam as shown in Figure~\ref{fig:2}a, the longer time dynamics shown in Supplementary Figure~9 and the partial recovery cannot be explained by this. The longer-lasting transient SHG suppression at higher fluences further indicates that the SHG responses are not solely dependent on pulse duration but reflect a genuine structural response encoded in these curves. To clarify this further, Figure~\ref{fig:4}b shows the results of phase-sensitive SHG measurements taken at an excitation fluence of approximately 400~mJ/cm$^2$, where the time-dependent SHG field interferes with a reference SHG pulse generated in a non-excited crystal. Shifts in the interference fringes indicate changes in the phase of the SHG signal from the driven LiNbO$_3$ crystal~\cite{Mankowsky2017,Shen2013}. Initially, the SHG shifts due to the electro-optic effect near delay zero. Near 200~fs time delay, the SHG interference fringe transiently vanishes and is followed by a sudden sign change, indicating a 180$^{\circ}$ phase flip of the SHG field. The reversed phase remains constant for approximately 50~fs before disappearing and then switching back to the initial value. These observations indicate a transient reversal of ferroelectric polarization shortly after zero delay, closely resembling the phase flip from resonant nonlinear phononic approaches~\cite{Mankowsky2017}. 

\section{Mechanistic Discussion and Technological Significance}

We further conducted comparative experiments on lithium iodate ($\alpha$-LiIO$_3$), a non-centrosymmetric but non-ferroelectric material~\cite{Nash1969}, to isolate signals beyond the nonlinear wave mixing effect. The absence of FSRS peak sign change and observation of minor SHG change starkly contrasts with the LiNbO$_3$ case, which can be rationalized by the non-ferroelectric nature of $\alpha$-LiIO$_3$ (see Supplementary Figure~7, 10, and Supplementary Note~8). We note that the Raman force model provides a simple explanation for the transient nature of the ferroelectric reversal. Figure~\ref{fig:4}c illustrates the multi-dimensional potential energy surface, showing two local minima corresponding to the two ferroelectric phases. The vectors connecting these minima are not aligned with the soft mode but rather represent a superposition of multiple modes, with the soft mode being dominant (details are included in Supplementary Note~9). Resonant excitation methods~\cite{Mankowsky2017,Henstridge2022}, which rely on nonlinear coupling to the soft mode, do not offer this capability, whereas in our case, the excitation wavelength was not optimized for precise phase control between different modes. Selecting an optimal wavelength could align the nonresonant Raman force between the two phases, facilitating a metastable transition to the other ferroelectric phase.

To evaluate its technological significance, we analyzed the switched sample thickness and energy consumption in the non-resonant Raman excitation regime and compared it with other established or emerging phase switching methods based on resonant phononic or above-bandgap excitation. We gauged the transmission of MIR energy through the sample used during the condition where the ferroelectric switching can be measured, providing an upper limit of absorption as reflection is not accounted for in this case. By combining this data with the thickness of the sample and the switched volume estimated from the laser spot size, we estimate the switching heat load shown in Supplementary Table~1, \textit{i.e.}, energy applied divided by the volume of the material switched, to be $\sim$~5$\times$10$^{-2}$~aJ/nm$^3$ for LiNbO$_3$, which are among the lowest values for lattice-based phase switches~\cite{Zalden2019,Sie2019,Li2019,Nova2019,Liu2012,Zhang2016,Stoica2019,Henstridge2022}. Importantly, our switching depth exceeds hundreds of microns, among the longest reported, in stark contrast to the nm-$\mu$m penetration depth of other resonant photoexcitation methods.

\section{Conclusion}

In summary, we have demonstrated that nonresonant Raman excitation of multiple coherent phonons can drive large-amplitude distortions, with their relative directions tuned, when using ultrashort pulses with low photon energy. We achieved nonresonant controls by inducing nonequilibrium ferroelectric switching dynamics in LiNbO$_3$ using MIR pulses well below the bandgap. By probing MIR-excited LiNbO$_3$ with FSRS and SHG with phase sensitivity, we observed compelling evidence of ultrafast ferroelectric switching comparable to that realized by resonant phonon excitation. That is evident by observing the flip in the FSRS peak sign, significant SHG suppression, and $\pi$-phase shift in SHG. Further comparative analysis reveals the merits of nonresonant Raman control of material phases with ultra-low switching energy consumption. Extending Raman from spectroscopic tools~\cite{Raman1928} to harness Raman force for material phase control has been a long-pursued goal~\cite{Yan1985,Fahy1994}. This effort offers a pathway to amplify Raman displacements over multiple modes and synchronize Raman mode displacements across the complex potential surface to modify the equilibrium positions on demand. The future implementation of long excitation pulses offers an extended time period of potential modification, including the possibility for thermodynamic forces to facilitate the theoretically predicted optomechanical-driven phase transitions~\cite{Zhou2020a,Zhou2018}. This hitherto unexplored strategy unlocks the potential for comprehensive manipulation of intricate phase diagrams with reduced energy consumption and surpasses traditional carrier or resonant excitation approaches.


\section{Experimental Section}
\threesubsection{Sample details} The $x$-cut single-domain LiNbO$_3$ crystal with a thickness of 500~$\mu$m was purchased from MTI Corp. The LiIO$_3$ single crystal in the $\alpha$ phase with a thickness of 500~$\mu$m was obtained from United Crystals.\\
\threesubsection{Ultrafast MIR pulse generation} MIR pulses were produced using a 0.5 mm thick GaSe crystal by performing difference-frequency mixing on the signal and idler outputs of a high-energy optical parametric amplifier (OPA). The OPA was powered by 35 fs pulses from a commercial Ti:Sapphire regenerative amplifier with a central wavelength of 800 nm and a power output of 6 mJ. The signal and idler outputs of the OPA were measured at 1.2 mJ and 1 mJ, respectively. The MIR pulses produced by the GaSe crystal were initially filtered using MIR filters (Thorlabs FB5000-500 and Edmund Optics \#62-643). These pulses were then collected and magnified using a pair of 2-inch diameter parabolic mirrors before being tightly focused onto the sample by a third parabolic mirror with a 2-inch diameter and a 2-inch effective focal length. The strength of the MIR field was adjusted using a half-wave plate and polarizer (Thorlabs WP25H-K).\\
\threesubsection{FSRS probe} The FSRS setup schematic is shown in Supplementary Figure~1. The narrowband pulse with a center wavelength of approximately 800 nm was generated through spectral filtering using a grating. To generate the femtosecond continuum probe pulse, the signal beam at about 1.35 $\mu$m from the OPA was focused on a sapphire crystal. When these two pulses interacted simultaneously on the sample, it resulted in sharp vibrational gain features appearing on top of the probe envelope. The white light pulse transmitted through the sample was directed towards a fast spectrometer (Ocean Insight, OCEAN-FX-VIS-NIR), capable of taking the spectrum at the same speed as the laser repetition rate of 1 kHz. To suppress intensity fluctuations in the MIR pulse, two synchronized differential choppers were used, operating at one-half and one-quarter of the laser repetition rate. The pulse sequence produced simultaneous results, including the probe background, ground- and excited-state FSRS, transient absorption spectra, and their differential FSRS spectrum~\cite{Dietze2016,Kloz2012}.\\
\threesubsection{SHG probe with phase sensitivity} The schematic illustration of the phase-sensitive SHG setup is provided in Supplementary Figure~2, and we adopted a near-collinear pump-probe geometry used in our experiments. For the phase-sensitive SHG measurements, we measured the phase of the emitted SHG electric field from the sample at all time delays by interfering it with the SHG light generated in a second LiNbO$_3$ crystal. The interference pattern was imaged on a CCD camera (Andor Luca). Detailed data analysis of phase-sensitive SHG measurements is included in Supplementary Note~7.\\
\threesubsection{FSRS simulation details of LiNbO$_3$} We first discuss the stimulated Raman scattering case in the absence of the MIR pump followed by the case with MIR-driven Raman excitation. For FSRS, We first solve the damped oscillator equation to obtain the Raman mode driven by FSRS pump and probe fields
\begin{equation} \label{eq:damped_ode_simplified}
	\begin{aligned}
		\ddot{Q}(z,t)+2\gamma\dot{Q}(z,t)+\omega_0^2 Q(z,t)&=\frac{\alpha_0'E_{n}(z,t) E_{c}(z,t)}{m}\\
	\end{aligned}
\end{equation}
where $Q$ is the coordinate of the phonon mode, $\gamma$ is the damping coefficient, $\omega_0$ is the inherent oscillation frequency of the molecule, $m$ is the reduced mass of the oscillator, $\alpha_0'$ is the first order derivative of polarizability $\alpha_0$ with respect to the coordinate $Q$, $E_n$ is the electric field of the narrowband Raman pump pulse, and $E_c$ is the electric field of the white-light continuum pulse. Then, we substitute the nonlinear polarization $P^{NL}(z,t)$ given below into the wave equation.
\begin{equation} \label{eq:PNL_t}
	\begin{aligned}
		P^{NL}(z,t)&=N\frac{\partial \alpha_0}{\partial Q}Q(z,t)[E_{n}(z,t)+E_{c}(z,t)]\\
		&\approx N\alpha_0'Q(z,t)E_{n}(z,t)
	\end{aligned}
\end{equation}
After approximations (see Supplementary Note~6), we integrate over the $z$ component in the wave equation and obtain the corresponding electric field in the frequency domain:
\begin{equation} \label{eq:P_to_E_simp}
	\begin{aligned}
		E(\omega)&=\frac{i\omega lN\alpha_0'}{2c\epsilon_0 n}FT\{ Q(t)E_{n}(t)\}\\
	\end{aligned}
\end{equation}
where $l$ is the sample thickness, $n$ is the sample refractive index, $FT\{\}$ refers to taking Fourier transform on the function in the bracket. The FSRS transmission is given by
\begin{equation} \label{eq:FSRS_trans}
	\begin{aligned}
		FSRS\; transmission=\frac{|E_{c}(\omega)+E(\omega)|^2}{|E_{c}(\omega)|^2}\\
	\end{aligned}
\end{equation}
where $E_{c}(\omega)=FT\{E_{c}(t)\}$. Upon excitation with the MIR pulse, additional contributions to the FSRS process arise from parasitic nonlinear mixing processes. Here, the MIR-excited phonon modes are not modulating the phonon modes excited by the FSRS process. 
Here, the MIR pulse irradiates the material and excites a given phonon mode $Q_{MIR}(t)$. Due to the presence of $Q_{MIR}(t)$, when the Raman pump pulse $E_{n}$ and Raman probe pulse $E_{c}$ arrive, the fields $E_{n}'$ and $E_{c}'$ are induced:
\begin{equation} \label{eq:pump_prime_probe_prime}
	\begin{aligned}
		E_{n}' = \frac{\partial P_{n}^{NL}(t)}{\partial t} = \partial (N\alpha_{0,{MIR}}'E_{n}(t)Q_{MIR}(t))/\partial t\\
		E_{c}' = \frac{\partial P_{c}^{NL}(t)}{\partial t} = \partial (N\alpha_{0,{MIR}}'E_{c}(t)Q_{MIR}(t))/\partial t\\
	\end{aligned}
\end{equation}
where $\alpha_{0,{MIR}}'$ is the polarizability of the vibrational mode $Q_{MIR}$. The field $E_{n}'$ generates a six-wave mixing term in the nonlinear polarization via mixing with the mode $Q$ excited in equation \eqref{eq:damped_ode_simplified}, and the field $E_{c}'$ imparts a delay-dependent frequency shift on the FSRS peak corresponding to mode $Q$. The modeling process used to calculate these effects on the time-resolved FSRS spectra is described in Supplementary Note~6.


In our simulations, we model the time-dependent response of $\alpha_0'$ as
\begin{equation} \label{eq:pump_prime_probe_prime}
	\begin{aligned}
		\alpha_0'(t)=(1&-A\frac{1}{2\sqrt{2\pi}}e^{\frac{-(t+2\tau)^2}{4\tau^2}}(1+erf(\alpha \frac{t+\tau}{\sqrt{2}\tau})\\
		&+B(\frac{1}{2}erf(\frac{t}{\tau})+\frac{1}{2})e^{-\frac{t}{5\tau}}\cos(2\pi f_1 t+\phi_1)\\
		&+C(\frac{1}{2}erf(\frac{t}{\tau})+\frac{1}{2})e^{-\frac{t}{5\tau}}\cos(2\pi f_2 t+\phi_2))\alpha_0'(+\infty)\\
	\end{aligned}
\end{equation}
Here $t=0$ corresponds to the time when the MIR pulse irradiates the sample. We choose $A=5$, $B=0.03$, $C=0.5$, $\alpha=1$~A$^2$s$^4$kg$^{-1}$, $f_1=8\times 10^{12}$~Hz, $f_2=1\times 10^{12}$~Hz, $\phi_1=0$, $\phi_2=-\frac{2\pi}{3}$, $\tau=1\times 10^{-13}$~s. \\

\medskip
\textbf{Supporting Information} \par 
Supporting Information is available from the Wiley Online Library or from the author.

\medskip
\textbf{Acknowledgements} \par 
This work was supported by the US Department of Energy (DOE), Office of Basic Energy Sciences, Division of Materials Sciences and Engineering, under contract no. DE-AC02-76SF00515. I.S., J.S. acknowledge the start-up funding provided by the University of Washington. H.X., J.L. acknowledge support by an Office of Naval Research MURI through grant \#N00014-17-1-2661. Work at UT Austin was supported by the National Science Foundation under grant DMR-2308817 (to X.P. for data taking), the Air Force Office of Scientific Research under Young Investigator Program award FA9550-24-1-0097 (to S.Z. for data taking), and the ARL-UT Austin Cooperative Agreement W911NF-21-2-0185 (to E.B. for data interpretation and supervision). J.Z. acknowledges the support by the National Natural Science Foundation of China under grant No. 21903063. C.H. acknowledges support from the U.S. Department of Energy, Office of Science through the AMOS program and the Alexander von Humboldt Research Fellowship. S.G. acknowledges support from the U.S. Department of Energy, Office of Science through the AMOS program. K.A.N. acknowledges support from the U.S. Department of Energy, Office of Basic Energy Sciences, under Award No. DE-SC0019126.

\textbf{Conflict of Interest} \par 
The authors declare no conflict of interest.

\textbf{Author Contributions} \par 
J.S., C.H., H.X., and Y.S. contributed equally. J.S. and A.M.L. conceived the idea. J.S., Y.S., and A.M. performed the measurements, data analysis, and simulation under the guidance of A.M.L. I.S. performed the measurements under the supervision of J.S. C.H. performed the measurements and data analysis under the guidance of D.A.R. and S.G. M.H. performed the FSRS modeling. X.P. and S.Z. performed the spontaneous Raman scattering measurements, under the supervision of E.B. H.X., J.Z., and J.L. performed the theoretical simulation. M.H., M.T., T.F.H., K.A.N., E.B., and D.A.R. contributed to the data interpretation. 

\textbf{Data Availability Statement} \par 
The data that support the findings of this study are available from the corresponding author upon reasonable request.

\medskip

%
\bibliographystyle{MSP}
\providecommand{\noopsort}[1]{}\providecommand{\singleletter}[1]{#1}%

\clearpage
\newpage

\begin{figure}
  \includegraphics[width=\linewidth]{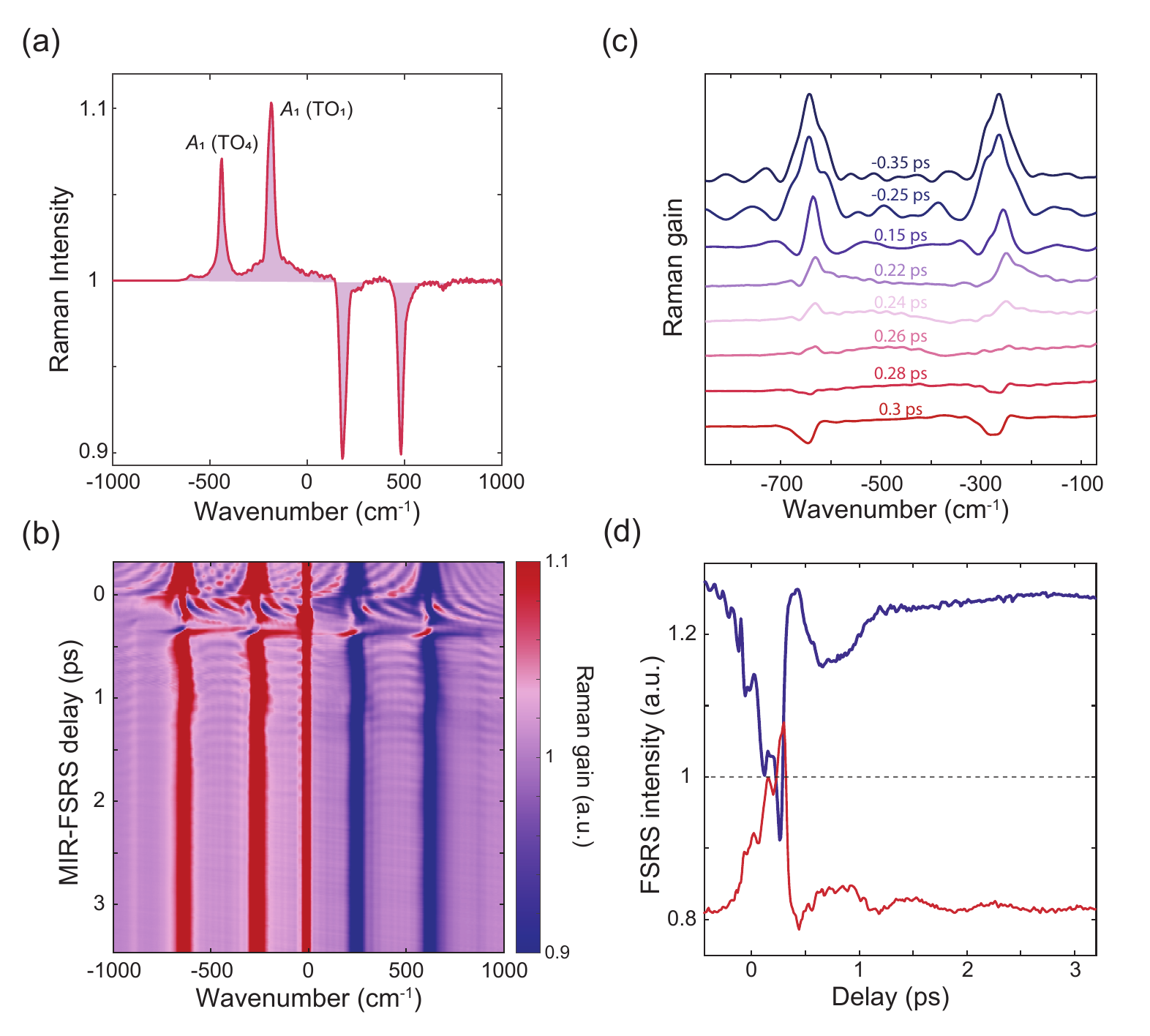}
  \caption{MIR-driven ferroelectric reversal in FSRS. a) The FSRS spectrum for LiNbO$_3$ at equilibrium shows two peaks corresponding to the $A_1$ (TO$_1$) and $A_1$ (TO$_4$) modes. b) The evolution of FSRS peaks as a function of MIR-FSRS delays. Besides the periodic modulation of FSRS peak positions occurring after the delay $\sim$~350~fs, the intensity of the FSRS peaks is dramatically suppressed and exhibits a sign reversal at a MIR-FSRS delay of approximately 300~fs. c) The horizontal line cut of \textbf{b} shows the FSRS spectrum at various MIR-FSRS delay. d) The vertical line cut of \textbf{b} at $\sim$~600 and -600 cm$^{-1}$ shows the time-dependent FSRS peak intensities as a function of MIR-FSRS delay. A transient sign reversal followed by intensity modulation by phonon coherence is evident.}
  \label{fig:3}
\end{figure}

\begin{figure}
	\includegraphics[width=\linewidth]{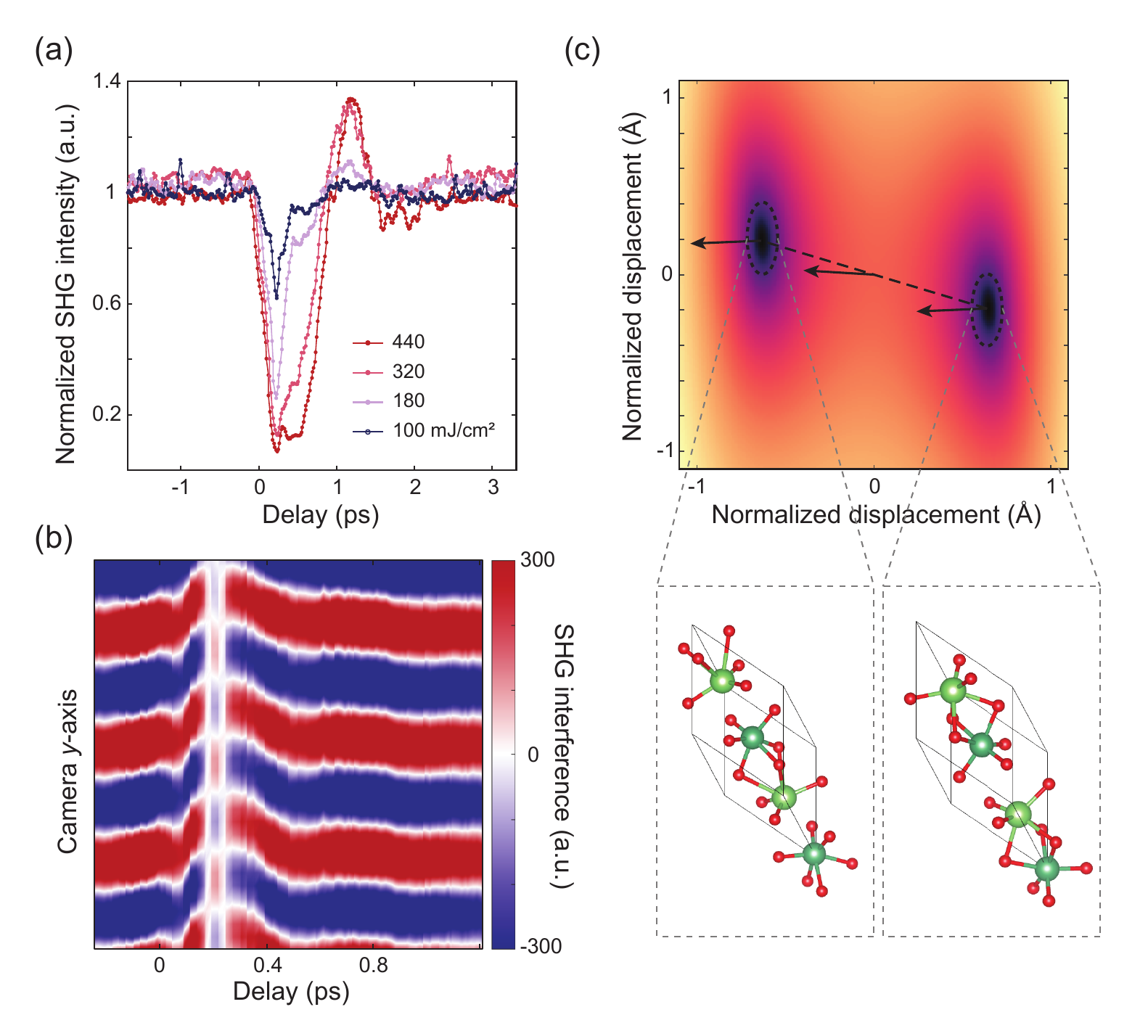}
	\caption{MIR-driven ferroelectric reversal in SHG and explanation of its transient nature. a) Time-resolved SHG intensity in LiNbO$_3$ normalized to its value before excitation at varying excitation fluences. Following MIR excitation at a fluence of $\sim$~440~mJ/cm$^2$ close to its damage threshold, the SHG intensity reduces substantially, followed by transient recovery before relaxing back. The SHG response is not solely dependent on pulse duration, as evidenced by the significant broadening of transient suppression with increasing fluences. Full fluence-dependent data can be found in Supplementary Figure~9. b) Time-resolved measurement of the SHG interference fringes with MIR excitation at a fluence of $\sim$~400~mJ/cm$^2$. Besides the dynamic shifting in the fringe positions linked to the phonon coherence, the data shows a transient $\pi$ phase change (sign reversal of the SHG fringe) at about delay 200~fs. c) Two-dimensional potential energy plot as a function of normalized displacements along two phonon coordinates in LiNbO$_3$. The displacement between two ferroelectric phases (bottom plots) can approximate the phonon normal modes at the local minimum and the saddle point, marked by the arrow. They represent different modes and do not align with the displacement vector connecting the two ferroelectric phases.}
	\label{fig:4}
\end{figure}

\end{document}